\documentclass[longauth]{aa}  
\usepackage{ulem}
\usepackage{graphicx}
\usepackage{txfonts}
\usepackage{xcolor}
\usepackage[colorlinks=true,linkcolor=red,anchorcolor=blue,citecolor=blue,filecolor=black,menucolor=black,runcolor=black,urlcolor=blue]{hyperref}

\newcommand{\OIII}[1]{[\ion{O}{III}]\,#1}
\newcommand{\SIII}[1]{[\ion{S}{III}]\,#1}
\newcommand{\Ha}{H$\alpha$\xspace}
\newcommand{\Hb}{H$\beta$\xspace}

\newcommand{\Mstar}{$M_\mathrm{*}$\xspace}
\newcommand{\Msyr}{\,$M_\odot$\,yr$^{-1}$\xspace}

\newcommand{\arcs}{$\arcsec$\xspace}
\newcommand{\micron}{\,$\mu$m\xspace}
\newcommand{\Myr}{\,Myr\xspace}

\newcommand{\Msun}{\,$M_\mathrm{\odot}$\xspace}

\newcommand{\Zsun}{\,$Z_\mathrm{\odot}$\xspace}

\newcommand{\Angs}{$\AA$\xspace}

\newcommand{\kms}{\,km\,s$^{-1}$\xspace}

\newcommand{\uJy}{\,$\mu$Jy\xspace}

\newcommand{\nJy}{\,nJy\xspace}
\newcommand{\ergs}{\,erg\,s$^{-1}$\xspace}
\newcommand{\ergscm}{\,erg\,s$^{-1}$\,cm$^{-2}$\xspace}

\newcommand{\EXP}[1]{\,$\times$\,10$^{#1}$}
\newcommand{\range}{\,$-$\,}
\newcommand{\eg}{e.g.\,}
\newcommand{\ie}{i.e.\,}
\newcommand{\eqq}{\,=\,}
\newcommand{\pmm}{\,$\pm$\,}
\newcommand{\simi}{$\sim$\,}
\newcommand{\z}{$z$\xspace}

\begin{document}

   \title{MIRI spectrophotometry of GN-z11: Detection and nature of an optical red continuum component}

   \author{A. Crespo G\'omez\inst{\ref{inst:STScI}}
    \and L. Colina \inst{\ref{inst:CAB}}
    \and P. G. P\'erez-Gonz\'alez \inst{\ref{inst:CAB}}
    \and J. \'Alvarez-M\'arquez \inst{\ref{inst:CAB}}
    \and M. Garc\'ia-Mar\'in\inst{\ref{inst:ESA}}
    \and A. Alonso-Herrero\inst{\ref{inst:CAB-ESAC}}
    \and M. Annunziatella\inst{\ref{inst:CAB}} 
    \and A. Bik\inst{\ref{inst:Stockholm}}
    \and S. Bosman\inst{\ref{inst:Heidelberg},\ref{inst:MPIA}}
    \and A. J. Bunker\inst{\ref{inst:Oxford}}
    \and A. Labiano\inst{\ref{inst:CAB-ESAC},\ref{inst:Telespazio}} 
    \and D. Langeroodi\inst{\ref{inst:DARK}} 
    \and P. Rinaldi\inst{\ref{inst:Steward}}
    \and G. {\"O}stlin\inst{\ref{inst:Stockholm}} 
    \and L. Boogaard\inst{\ref{inst:Leiden}} 
    \and S. Gillman\inst{\ref{inst:DTU}, \ref{inst:DAWN}}
    \and G. Barro\inst{\ref{inst:Stockton}}
    \and S. L. Finkelstein\inst{\ref{inst:Texas}}
    \and G. C. K. Leung \inst{\ref{inst:MIT}}
   }
   
   \institute{Space Telescope Science Institute (STScI), 3700 San martin Drive, Baltimore, MD 21218, USA\\    \email{acrespo@stsci.edu} \label{inst:STScI}
    \and Centro de Astrobiolog\'{\i}a (CAB), CSIC-INTA, Ctra. de Ajalvir km 4, Torrej\'on de Ardoz, E-28850, Madrid, Spain \label{inst:CAB}
    \and European Space Agency, Space Telescope Science Institute, Baltimore, Maryland, USA \label{inst:ESA} 
    \and Centro de Astrobiolog\'ia (CAB), CSIC-INTA, Camino Viejo del Castillo s/n, 28692 Villanueva de la Ca\~{n}ada, Madrid, Spain \label{inst:CAB-ESAC}
    \and Department of Astronomy, Stockholm University, Oscar Klein Centre, AlbaNova University Centre, 106 91 Stockholm, Sweden \label{inst:Stockholm}
    \and Institute for Theoretical Physics, Heidelberg University, Philosophenweg 12, 69120 Heidelberg, Germany \label{inst:Heidelberg}
    \and Max-Planck-Institut f\"ur Astronomie, K\"onigstuhl 17, 69117 Heidelberg, Germany\label{inst:MPIA}
    \and Department of Physics, University of Oxford, Denys Wilkinson Building, Keble Road, Oxford OX13RH, U.K. \label{inst:Oxford}
    \and Telespazio UK for the European Space Agency, ESAC, Camino Bajo del Castillo s/n, 28692 Villanueva de la Ca\~{n}ada, Spain \label{inst:Telespazio}
    \and DARK, Niels Bohr Institute, University of Copenhagen, Jagtvej 128, 2200 Copenhagen, Denmark \label{inst:DARK}
    \and Steward Observatory, University of Arizona, 933 North Cherry Avenue, Tucson, AZ 85721, USA \label{inst:Steward}
    \and Cosmic Dawn Center, DTU Space, Technical University of Denmark, Elektrovej 327, 2800 Kgs. Lyngby, Denmark \label{inst:DTU}
    \and Cosmic Dawn Centre, Copenhagen, Denmark \label{inst:DAWN} 
    \and University of the Pacific, Stockton, CA 90340, USA\label{inst:Stockton}
    \and Leiden Observatory, Leiden University, PO Box 9513, NL-2300 RA Leiden, The Netherlands\label{inst:Leiden}
    \and Department of Astronomy, The University of Texas at Austin, Austin, TX, USA \label{inst:Texas}
    \and MIT Kavli Institute for Astrophysics and Space Research, 77 Massachusetts Ave., Cambridge, MA 02139, USA \label{inst:MIT}
    }

  \abstract
{We present new MIRI F560W, F770W, and F1000W imaging of the galaxy GN-z11 at a redshift of 10.603. We report a significant detection (14$\sigma$) in the F560W and F770W images, and a marginal detection (3.2$\sigma$) in the F1000W filter. The new MIRI observations cover the optical-red spectral range and significantly extend previous NIRCam wavelength coverage from rest-frame 0.38\micron up to 0.86\micron. In this work, we analyse the spectral energy distribution (SED) combining this new MIRI imaging data with archival NIRSpec/Prism and MRS spectroscopy, and NIRCam imaging, i.e. covering the rest-frame 0.12\range0.86\micron. New constraints such as the equivalent widths of the strong optical lines (\OIII{$\lambda$5008}, H$\beta$ and H$\alpha$) and the continuum emission at rest-frame 0.48$\mu$m, 0.66$\mu$m, and 0.86$\mu$m, free of emission line contributions, are presented. The continuum emission shows a flat energy distribution, in $f_\nu$, up to 0.5 $\mu$m, compatible with the presence of a mixed stellar population of young (4\pmm1\,Myr) and mature (63\pmm23\,Myr) stars that also account for the [\ion{O}{III}], \Hb, and \Ha emission lines. The continuum at rest-frame 0.66\micron shows a 36\pmm3\% flux excess above the predicted flux for a mixed stellar population, pointing to the presence of an additional source contributing at these wavelengths. This excess increases to 91\pmm28\% at rest-frame 0.86\micron, although with a large uncertainty due to the marginal detection in the F1000W filter.  
We consider that hot dust emission in the dusty torus around a type 2 active galactic nucleus (AGN) could be responsible for the observed excess. Alternatively, this excess could be due to hot dust emission or a photoluminiscence dust process (Extended Red Emission, ERE) under the extreme UV radiation field, as is observed in local metal-poor galaxies and in young compact starbursts. The presence of a type 1 AGN is not supported by the observed SED as the hot dust emission in luminous high-\z quasi-stellar objects (QSOs) contributes at wavelengths above rest-frame 1\micron, and an additional ad hoc red source would be required to explain the observed flux excess at 0.66 and 0.86\micron. Additional deep MIRI imaging covering the rest-frame near-IR is needed to confirm the flux detection at 10\micron, and to discriminate between the different hot dust emission in the extreme starburst and AGN scenarios.}

   \keywords{galaxies: high-redshift --galaxies: individual: GN-z11 -- galaxies: starburst -- infrared: galaxies }

   \maketitle

%-------------------------------------------------------------------

\section{Introduction}
\label{sec:intro}

GN-z11 was firstly identified as a \z$>$\,10 Lyman-break galaxy candidate using multi-colour imaging from HST CANDELS \citep{Bouwens+10}. A subsequent HST grism spectroscopic analysis revealed a flux dropout identified as the Lyman break by \citet{Oesch+16}, yielding a redshift value of \z\simi11.09. The arrival of JWST has enabled the study of this object at optical and near-IR wavelength with unprecedented sensitivity and spatial resolution. In particular, the JWST Advanced Deep Extragalactic Survey (JADES; \citealt{Eisenstein+23}) derived a redshift of \z{}\eqq10.6034\pmm0.0013 using multiple emission lines based on NIRSpec/Prism data \citep{Bunker+23}. Besides, this NIRSpec spectrum has revealed the presence of a high ionisation field (log($U$)\,\simi$-$2.3) and a super-solar abundance of nitrogen ([N/O]$>$\,$-$0.49). Recent JWST NIRCam imaging has shown that GN-z11 UV emission can be depicted as the combination of a point-source and an extended component, showing an extremely compact morphology (combined $R_\mathrm{eff}$ of 64\pmm20\,pc) and large UV luminosity ($M_\mathrm{UV}$\eqq$-$21.58\pmm0.02; \citealt{Tacchella+23_GNz11}).

JWST multi-wavelength photometry and spectroscopy result in star formation rates (SFRs) and stellar masses of between 20\range30\Msyr and log(\Mstar/$M_{\odot}$)\eqq8.7$-$9.1 \citep{Bunker+23,Tacchella+23_GNz11}, respectively. The relatively high stellar mass at only \simi420 Myr after the Big Bang suggests a rapid build-up of stellar mass, which has been also found in other JWST spectroscopically confirmed galaxies at \z\,$>$\,10 \citep{Robertson+23,Curtis-Lake+23,Carniani+24,Naidu+25}. The exceptionally UV brightness of this object allows its observation using ground-based facilities. Recent NOEMA observations \citep{Fudamoto2024} have determined upper limits on the [\ion{C}{II}]158\micron emission and 160\micron dust continuum yielding log($M_\mathrm{mol,\,[\ion{C}{II}]}$/$M_{\odot}$)\,$<$\,9.3 and log($M_\mathrm{dust}$/$M_{\odot}$)\,$<$\,6.9. These results are consistent with a negligible dust attenuation and the blue colour ($\beta$\eqq-2.4) obtained from a spectral energy distribution (SED) fitting analysis \citep{Tacchella+23_GNz11}.

Using medium-resolution (R\eqq1000) NIRSpec observations, \cite{Maiolino+24_BH} present several pieces of evidence favouring the scenario in which GN-z11 is a narrow-line Seyfert 1 (NLS1) with a black hole mass of log($M_{\mathrm{BH}}/M_{\odot}$)\eqq6.2\pmm0.3 accreting material at a large Eddington ratio ($\lambda_\mathrm{Edd}$\eqq$L_\mathrm{bol}/L_\mathrm{Edd}$\simi5.5). The authors found the presence of high-excitation ($>$\,60\,eV) emission lines in the UV spectrum (\eg[\ion{Ne}{IV}]$\lambda\lambda$2422,\,2244) that are usually present in active galactic nuclei (AGNs) \citep{LeFevre+19,Terao+22}. In addition, the high critical density ($n_\mathrm{e,\,\ion{N}{III}]}$\,$>$\,10$^{5}$\,cm$^{-3}$) and broad (FWHM$_\mathrm{\ion{He}{II}\lambda1640}$\simi1200\kms) emission lines revealed by the NIRSpec data are typically observed in the broad-line regions (BLRs) of AGNs. The presence of an outflow powered by this AGN was also inferred from a high equivalent width (EW) and blueshifted absorption of a \ion{C}{IV} doublet \citep{Maiolino+24_BH}. The small and compact size of the BLR linked to this potential AGN would justify the large [N/O] of GN-z11 through a very rapid chemical enrichment driven by only a few supernovae (SNe). In addition, a blue bump
has been observed in the UV using NIRSpec/Prism data \citep{Ji+25}. This bump has been proposed to be linked to dense \ion{Fe}{II}-emitting clouds ($n_\mathrm{H}$\simi10$^{11}$\,cm$^{-3}$) infalling at \simi3000\kms. GN-z11 hosting an AGN and the presence of tentative nearby companions detected as Ly$\alpha$ blobs make this region a possible candidate for being the core of a protocluster with a halo mass of log($M_\mathrm{halo}$/$M_{\odot}$)\eqq10.4 \citep{Scholtz+24}.

However, alternative scenarios have been proposed to justify the overabundance of nitrogen in GN-z11, such as the presence of Wolf-Rayet (WR) stars \citep{Senchyna2024} or super-massive stars in a proto-globular cluster \citep{Charbonnel2023}. In addition, \cite{Bhatt2024} show that cosmological simulations do not support a super-Eddington type 1 AGN scenario for GN-z11. The analysis of the \Ha and \OIII{$\lambda$5008} lines observed with MIRI/MRS carried out by \cite{Alvarez-marquez+25} does not support the presence of an accreting black hole dominating the optical emission lines and continuum in GN-z11. In the case in which \Ha and the optical continuum are entirely produced by the BLR, the accretion rate and bolometric luminosity would be unrealistically large, but the black hole mass would be smaller (log($M_\mathrm{BH}$/$M_{\odot}$)\eqq5.77) than the one inferred by UV lines in \cite{Maiolino+24_BH}. In addition, $L_\mathrm{H\alpha}/L_\mathrm{X,2-10\,keV}$ and $L_\mathrm{H\alpha}$/$L_\mathrm{5100}$ were found to be inconsistent with the low-\z AGN relations \citep{Ho2001,Greene+05b}.

To date, the majority of studies on GN-z11 have relied on UV data. However, we still lack important information at optical wavelengths, where the contributions from the young stellar populations begin to decrease and those from more evolved populations, or redder emission sources, are expected to become more prominent. So far, only \citet{Alvarez-marquez+25} have covered the optical wavelengths, using MIRI/MRS data, detecting \OIII{$\lambda$5008} and \Ha emission consistent with star formation. This paper presents the first mid-infrared F560W and F770W images of GN-z11, tracing its optical emission at rest-frame 0.5 and 0.66\micron, respectively. We combine these images with ancillary spectro-photometric data covering the rest-frame 0.15 up to 0.86\micron to trace the UV to optical emission continuum of this object and extend the analysis of its SED up to optical-red wavelengths. The paper is organised as follows. Section~\ref{sec:General_observations} presents the new JWST/MIRI images used in this work and the ancillary JWST data. Section~\ref{sec:Analysis} describes the continuum-only extraction and SED fitting analysis carried out. The main results drawn from the analysis and their discussion are presented in Section~\ref{sec:Results}. Finally, Section~\ref{sec:Summary} summarises the main conclusions and results of this work. Throughout this paper we assume a Chabrier initial mass function (IMF, \citealt{Chabrier+03}) and a flat $\Lambda$ cold dark matter ($\Lambda$CDM) cosmology, with $\Omega_\mathrm{m}$\,=\,0.31 and H$_0$\,=\,67.7\,km\,s$^{-1}$\,Mpc$^{-1}$ \citep{PlanckCollaboration18VI}. For this cosmology, 1 arcsec corresponds to 4.08\,kpc at \z{}\eqq10.603, while the luminosity distance is $D_\mathrm{L}$\eqq113.4\,Gpc.

\section{Observation and data processing}
\label{sec:General_observations}

\subsection{JWST/MIRI data and calibration}
\label{subsec:MIRIdata}

The JWST images of the GN-z11 were obtained on December 8, 2023 using the  MIRI imager \citep[MIRIM,][]{Bouchet+15} with the F560W and F770W filters, as part of the cycle 1 JWST program ID\,1264 (PI: Colina, L.). The observations were performed with the FASTR1 read-out mode in a five- and seven-point dither, for F560W and F770W, respectively, with a medium-size cycling pattern. The total on-source integration time corresponds to 2012 and 5866 seconds distributed in 10 and 28 integrations, respectively, for F560W and F770W.

These MIRIM images have been calibrated using the JWST pipeline (v1.12.0) with the context 1170 of the Calibration Reference Data System (CRDS). This CRDS includes photometric calibrations considering the temporal evolution and aperture corrections taking into account the point spread function (PSF) cruciform. In addition to the general procedure, further steps have been applied to correct for striping and background gradients (see more details in \citealt{Alvarez-Marquez+23,Perez-Gonzalez+24,Ostlin+25}). The process concluded with the creation of final MIRI images dithered to a pixel scale of 0.06$\arcsec$/pixel. According to the JWST documentation\footnote{https://jwst-docs.stsci.edu/jwst-mid-infrared-instrument/miri-performance/miri-point-spread-functions}, the spatial resolutions for these images are FWHM\,=\,0.207\arcs and 0.269\arcs, for the F560W and F770W, respectively. Following this calibration, we obtained a 3$\sigma$ depth of 26.67 and 26.96\,mag for F560W and F770W for unresolved point sources, respectively.

\subsection{Ancillary JWST data: Photometry and spectroscopy}
\label{subsec:ancillary_jwst}

In addition to the MIRI F560W and F770W images, this work takes advantage of the large amount of high-quality JWST data available for GN-z11 covering a large wavelength range. To trace the near-IR counterpart of this galaxy, we used the F1000W and F2100W images available from the MEOW survey (PID\,5407, PI: G. Leung) which covers up to rest-frame \simi2\micron. These observations were performed with a total exposure time of 722 and 3086\,s reaching a 3$\sigma$ level of 230\nJy and 2.6\uJy (\ie 25.5 and 22.9\,mag, respectively). In addition, the 11 NIRCam images available from JADES DR3 \citep{Eisenstein+23,Bunker+24,DEugenio+25}, with median 3$\sigma$ depths of \simi5\nJy (\ie\simi29.6\,mag, \citealt{DEugenio+25}), were used to trace the UV continuum from rest-frame 0.1 up to \simi0.4\micron. We performed aperture photometry as is explained in Sect.~\ref{subsec:phot} for all the NIRCam and MIRI images available.

We also incorporated JWST ancillary spectroscopic data for this object. In particular, we considered the fluxes from the UV lines detected on the NIRSpec/Prism data presented in \cite{Bunker+23} after a re-normalisation to match the NIRCam photometry (see Sect.~\ref{subsec:phot} and App.~\ref{app:Phot_norm}). In addition to the UV lines from NIRSpec, we also used the flux from the optical lines present in the MRS data and extracted by \citet{Alvarez-marquez+25}. The MRS spectra contain the \Hb+\,[\ion{O}{III}] and \Ha lines that fall into the MIRI/F560W and F770W wavelength ranges, respectively. We did not apply any normalisation as the MRS line fluxes were corrected for aperture losses using the MRS PSFs \citep{Argyriou+23}.

\section{Analysis}
\label{sec:Analysis}

\subsection{Aperture photometry and NIRSpec re-normalisation}
\label{subsec:phot}

\begin{table}
\centering
\caption{NIRCam and MIRI photometry}
\begin{tabular}{c|c|c|c}
\hline
          Filter & units &  Obs. & Cont. level        \\
\hline 
NIRCam/F115W & nJy & 31\,$\pm$\,9 & - \\
NIRCam/F150W & nJy & 118\,$\pm$\,9 & - \\
NIRCam/F182M & nJy & 177\,$\pm$\,11 & 154\,$\pm$\,12\\
NIRCam/F200W & nJy & 201\,$\pm$\,7 & 185\,$\pm$\,8 \\
NIRCam/F210M & nJy & 175\,$\pm$\,14 & 164\,$\pm$\,14 \\
NIRCam/F277W & nJy & 161\,$\pm$\,3 & 161\,$\pm$\,3 \\
NIRCam/F335M & nJy & 130\,$\pm$\,7 & 123\,$\pm$\,7 \\
NIRCam/F356W & nJy & 153\,$\pm$\,3 & 150\,$\pm$\,3 \\
NIRCam/F410M & nJy & 133\,$\pm$\,6 & 129\,$\pm$\,6 \\
NIRCam/F444W & nJy & 163\,$\pm$\,4 & 140\,$\pm$\,4 \\
MIRIM/F560W & nJy & 368\,$\pm$\,26 & 156\,$\pm$\,31 \\
MIRIM/F770W & nJy & 274\,$\pm$\,20 & 212\,$\pm$\,22 \\
MIRIM/F1000W & nJy & 249\,$\pm$\,77 & 237\,$\pm$\,77$^\dagger$ \\
MIRIM/F2100W & $\mu$Jy & $<$2.6 & $<$2.6 \\
\end{tabular}
    \label{tab:fluxes}
    \tablefoot{The third and fourth columns display the observed and continuum level, i.e. the measured photometry and the continuum fluxes after subtracting the emission line contribution (see Sect~\ref{subsec:Cont-only}), respectively. We adopted 3$\sigma$ upper limits for F2100W. $^\dagger$: The continuum level for F1000W was derived assuming a \SIII{{}$\lambda\lambda$9071,9533}/\Ha{}\eqq0.17 (see Sect.~\ref{subsec:Cont.-results}).}
\end{table}

\begin{figure*}
\centering
   \includegraphics[width=\linewidth]{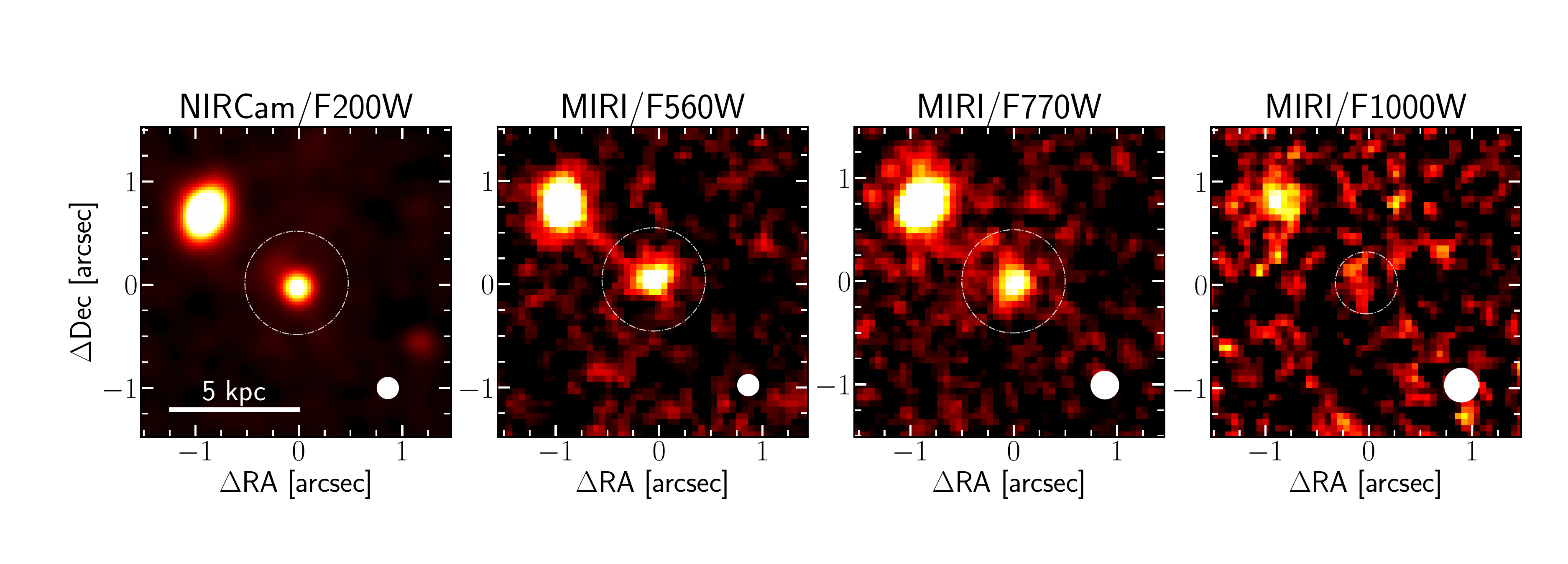}
      \caption{GN-z11 images in the NIRCam/F200W, MIRI/F560W, MIRI/F770W, and MIRI/F1000W filters. The NIRCam/F200W image was convolved to match the MIRI/F560W resolution (see Sect.~\ref{subsec:phot}). Dashed circles represent the 0.5\arcs aperture used for the photometry extraction. The white circles in the lower right corners of each panel represent the FWHM of each filter. }
         \label{fig:Aperture_plot}
\end{figure*}

We extracted the GN-z11 photometry from the available NIRCam and MIRI images considering circular apertures, centred on the F560W emission peak. The size of these apertures (\ie$r$\,=\,0.5\arcs) has been chosen to fully encompass both the UV and the optical emission. Before extracting the fluxes from NIRCam, we matched the PSF of these images to the spatial resolution of MIRI/F560W. At MIRI resolution, we consider GN-z11 to be a point-like source, and therefore we applied the aperture corrections derived from  the empirical PSFs presented in \citet{Libralato+24}. At $r$\,=\,0.5\arcs, these corrections are 1.29, 1.33, and 1.54 for MIRI/F560W and the NIRCam images, MIRI/F770W and MIRI/F2100W, respectively. To optimise the signal-to-noise (S/N) on the F1000W detection we assumed an aperture of $r$\,=\,0.3\arcs, applying an aperture correction factor of 1.71. The associated uncertainty of each photometric measurement was defined as the standard deviation of the local background around GN-z11 (1\arcsec<\,$r$\,<\,3\arcsec), masking all the additional objects present in the field of view (FoV). In this work, we have re-scaled the MIRI background noise by a factor of 2.24 to take into account the correlated noise induced when drizzling the individual observations (see \citealt{Ostlin+25}, for details). We adopted a 3$\sigma$ upper limit for the F2100W image, which does not present significant detection. Table~\ref{tab:fluxes} presents the flux values measured from the NIRCam and MIRI images. The S/N is generally high ($>$10) for all photometric values, with the exception of F1000W, in which GN-z11 is detected at a \simi3$\sigma$ level (see Fig.~\ref{fig:Aperture_plot}).

We also used the NIRCam fluxes to re-normalise the NIRSpec spectra, correcting from potential slit loses. In brief, we matched the NIRSpec/Prism spectra to the NIRCam photometry, multiplying by a factor of 1.325. A detailed description of the process is presented in Appendix~\ref{app:Phot_norm}.

\subsection{Continuum-only fluxes and EWs}
\label{subsec:Cont-only}

Since the optical emission lines detected in the MRS (\ie\OIII{$\lambda$5008}, \Ha) fall within the F560W and F770W filters, respectively, their contribution can be subtracted from the broadband fluxes, allowing us to estimate the underlying continuum at 5.6 and 7.7\micron and derive the EW of these emission lines. 

In this work, we have used the \OIII{$\lambda$5008} and \Ha fluxes presented by \citet{Alvarez-marquez+25} and the transmission curves for the F560W and F770W filters\footnote{Transmission curves are obtained from the \href{https://svo2.cab.inta-csic.es/theory/fps/}{SVO Filter Profile Service}}. For accuracy, we have included the \OIII{{}$\lambda$4960} and \Hb contributions using \OIII{$\lambda$5008}/\OIII{{}$\lambda$4960}\eqq3 and \Ha/\Hb{}\eqq2.86, considering case B and with negligible extinction \citep{Bunker+23,Alvarez-marquez+25}. Assuming log($U$)\eqq$-$3 and 0.2\Zsun, the potential contributions from [\ion{N}{II}]$\lambda\lambda$6549,\,6585 and [\ion{S}{II}]$\lambda\lambda$6718,\,6732 to the F770W flux are minimal ($<$5$\%$, \citealt{Curti+20}). 

In addition, we have extended the derivation of the continuum levels to UV wavelengths by using the NIRCam fluxes and the NIRSpec/Prism re-normalised emission lines presented in \citet{Bunker+23}. Finally, we estimated the continuum in MIRI/F1000W based on the \SIII{{}$\lambda\lambda$9071,9533}/\Ha{}\eqq0.17 ratio derived from \citet{Kewley+02}, assuming 0.2\Zsun and log($U$)\eqq$-$3. Although F1000W is slightly narrower than F770W, the \SIII{{}$\lambda\lambda$9071,9533} flux (\simi12\nJy) would contribute only up to \simi5$\%$ of the total F1000W photometry. The computed continuum-only fluxes for the NIRCam and MIRI filters are listed in Table~\ref{tab:fluxes}.

\citet{Alvarez-marquez+25} estimate an upper limit of a possible broad component in \Ha of 2\range 3\EXP{-18}\ergscm. If we consider this broad component, the F560W and F770W continuum levels would decrease by 6\range9\nJy and 18\range27\nJy, respectively, assuming a ratio \Ha/\Hb{}\eqq2.86 for the broad component. Based on the MIRI/F560W and F770W continuum-only fluxes, we derived rest-frame EW(\OIII{$\lambda$5008})\eqq786\pmm81\Angs and EW(\Ha)\eqq547\pmm72\Angs, considering only the narrow line contribution to \Ha.

%These \Hb+\OIII{4960,5008} and \Ha emission represent \simi58$\%$ and \simi23$\%${} of the F560W and F770W fluxes, respectively. After removing their contribution, the continuum fluxes at 5.6 and 7.7\micron is 156\pmm26 and 212\pmm20\,nJy, respectively. In \citet{Alvarez-marquez+25}, the authors estimate an upper limit of a possible broad component in \Ha of 2\range 3\EXP{-18}\ergscm. If we consider this broad component, the F770W continuum would decrease down to 185\range194\,nJy, while the F560W continuum level would be 145\range149\,nJy, assuming a ratio \Ha/\Hb\eqq2.86 for the broad component. From the continuum levels assuming only the presence of a narrow component, we derive rest-frame EW(\OIII{$\lambda$5008})\eqq786\pmm154\Angs and EW(\Ha)\eqq547\pmm90\Angs. 

\subsection{SED fitting}
\label{subsec:SED_fitting}

In order to investigate the nature of GN-z11, we analysed its SED based on the available spectro-photometric data. In particular, the SED constructed with the NIRCam+MIRI photometry and the NIRSpec+MRS spectroscopy was fitted with two codes, \texttt{CIGALE} v2025.0 \citep{Burgarella2005,Noll2009,Boquien2019} and \texttt{SYNTHESIZER-AGN} \citep{Perez-Gonzalez+03_Synthesizer,Perez-Gonzalez+08_Synthesizer}. Since the F1000W flux shows a lower significance compared to the other MIRI values, we excluded F1000W (and the upper limit for F2100W) from the SED fitting to prevent biasing the results. However, we extracted the modelled fluxes at these filters to evaluate their consistency.

The SED fitting from \texttt{CIGALE} makes use of the UV+optical emission lines and the MIRI fluxes. In addition, as the current version of \texttt{CIGALE} does not use spectroscopic data, we constructed pseudo-continuum fluxes by integrating the re-normalised NIRSpec/Prism spectrum in 750\Angs bins. The emission lines were assigned separate bins to avoid contaminating the pseudo-continuum values. These UV pseudo-continuums, which isolate the effect of the emission lines and trace the galaxy continuum emission, were considered during the SED fitting along with the MIRI fluxes. In our \texttt{CIGALE} model, the stellar component was fitted using the stellar population templates from the Binary Population and Spectral Synthesis library (BPASSv2.2, \citealt{Stanway+18}) with $Z$\eqq0.2\Zsun (following \citealt{Alvarez-marquez+25}) and a Chabrier IMF \citep{Chabrier+03} with an upper stellar mass limit of 100\Msun. Nebular emission was included via Cloudy models \citep{Ferland+17}, assuming $Z$\eqq0.2\Zsun, an electron density of 1000\,cm$^{-3}$, and an ionisation parameter of $\log(U)$\,=\,$-2.3$ (following \citealt{Bunker+23}). Dust extinction was assumed to follow the attenuation curve from \citet{Calzetti+00} with $A_\mathrm{V}$\,$<$\,0.3\,mag, constrained by the Balmer line ratios derived from previous NIRSpec and MRS analysis \citep{Bunker+23,Alvarez-marquez+25}. We also considered the potential presence of a non-stellar component (\ie type 1 or 2 AGN) by adding an AGN component modelled as described by \citet{Fritz+06}. This model assumes a broken power-law point source as the ionisation source and the presence of a smooth dusty torus. Since our dataset covers up to rest-frame 0.86\micron, we cannot fully constrain the full model parameter space. Therefore, for simplicity, in this work we fixed an opening angle of 60\degr and a $R_\mathrm{max}$/$R_\mathrm{min}$\eqq10, following the results found for high-\z luminous and massive quasi-stellar objects \citep[QSOs;][]{Bosman+25}, while we allowed the radial and angular indices to range between $\beta$\eqq[-1,\,-0.5] and $\gamma$\eqq[0,\,6], respectively. We tested both type 1 and type 2 scenarios by forcing the angle of the line of sight to the AGN polar axis to be $\theta$<20\degr{} and >70\degr, respectively.

For the SED fitting carried out with \texttt{SYNTHESIZER-AGN}, we used the re-normalised NIRSpec/Prism spectrum, the MRS optical lines, and the MIRI fluxes. Throughout the fitting, we considered the stellar population models described by \citet{Bruzual&Charlot+03}, hereafter BC03, assuming a \citet{Chabrier+03} IMF with stellar masses between 0.1 and 100\Msun. The nebular continuum and emission lines were modelled using Cloudy c23.01 \citep{Gunasekera+23} predictions for a variety of ionising SEDs (parametrised by a range of effective temperatures (8000\range120000\,K) and ionising photon flux ($10^{37}$\range10$^{60}$\,s$^{-1}$)), and a variety of electron densities (log($n_\mathrm{e}$/cm$^{-3}$)\eqq2\range4) and metallicities (0.02\range0.2\Zsun, with solar abundances for individual elements). The dust extinction is described by the \citet{Calzetti+00} attenuation law whereby we constrained the $A_\mathrm{V}$ to be lower than 0.30\,mag, as for the \texttt{CIGALE} fitting. The AGN emission was included following the QSO composite spectrum created by combining the UV+optical and near-IR templates from \citet{Selsing+16} (up to rest- 1.13\micron) and \citet{Glikman+06} (rest- 1.13\,$-$\,3.5\micron), respectively. This empirical template was added to the fitting procedure to represent the average accretion disc and dust emission in luminous type 1 AGNs. Additionally, we also considered the presence, without an energy balance, of an additional dusty torus to reproduce the observed red SED shape. This component was defined by using self-consistent AGN torus model from \citet{Polletta+06,Polletta+07}, which is a highly obscured ($A_\mathrm{V}$\eqq4) QSO with UV emission coming from scattered light and IR emission coming from hot and/or warm dust.

To quantify the dependence of the physical parameters when using different stellar population models, we compared the \texttt{CIGALE} results considering both the BC03 and BPASSv2.2 stellar templates. The BPASSv2.2 models predict stronger UV emission due to binary evolution, allowing stellar populations older than 10\,Myr to still produce significant ionising flux. This typically yields older ages, higher SFRs, and lower stellar masses, while BC03 favours younger and more massive solutions for GN-z11. However, these differences are $<$10\range15$\%$ and, in all cases, smaller than those introduced by assuming different star formation histories (SFHs). We also examined the potential effect of adopting a different extinction law during the SED fitting. For this comparison, we tested the \cite{Charlot&Fall2000} attenuation curve, in which the young stars and the nebular emission suffer an additional attenuation compared with the older stars ($>$\,10\Myr). We find that, in our case, the SED fitting using \cite{Charlot&Fall2000} lead to stellar masses only \simi0.1\,dex lower than those assuming \cite{Calzetti+00}, while variations in other physical parameters (e.g. SFR$_\mathrm{100\,Myr}$, $M_\odot$) are lower than 10$\%$. 
We therefore argue that neither the choice of the stellar population templates nor that of the attenuation curve affects any conclusion drawn from our analysis.

In an effort to discover the nature of GN-z11, in Sections~\ref{subsec:Onlystellar} and~\ref{subsec:2comp+AGN} we tested different SFHs and the possible presence of an AGN for both SED fitting codes. Differences in the SFHs and AGN modelling are presented and discussed in the following sections.

\section{Results and discussion}
\label{sec:Results}

\subsection{Continuum-only fluxes and colour-colour diagrams}
\label{subsec:Cont.-results}

\begin{figure*}
\centering
   \includegraphics[width=\linewidth]{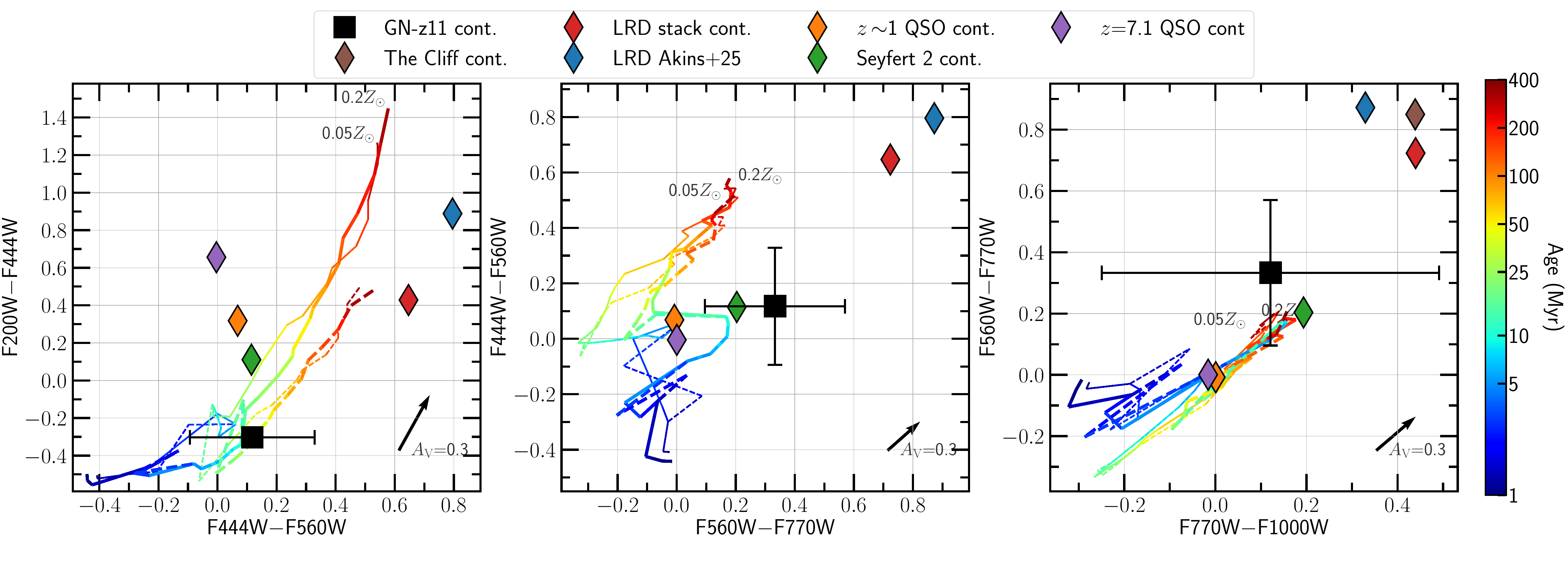}
      \caption{Continuum-only colour-colour diagrams. Black squares represent the GN-z11 colours based on the NIRCam and MIRI photometry after the subtraction of the emission lines contribution measured with NIRSpec and MRS (\ie continuum-only colours) presented in Table~\ref{tab:fluxes}. Continuum (dashed) lines display the continuum-only colours derived considering a stellar population colour-coded by their age assuming an instantaneous (continuum) SFH. These colours include the contribution from the stellar continuum calculated using the stellar models from BPASSv2.2 and its associated nebular continuum derived with Cloudy (see Sect.~\ref{subsec:Cont-only}). Thin and thick lines represent the 0.05 and 0.2\Zsun tracks, respectively. Orange, green, and purple diamonds show the derived continuum-only colours from the composed \z\simi1 QSO (\citealt{Glikman+06,Selsing+16}, see Sect.~\ref{subsec:SED_fitting}), the Seyfert 2 Mrk3 \citep{Spinelli+06}, and a \z{}\eqq7.1 QSO \citep{Bosman2024,Bosman+25}. Red, blue, and brown diamonds represent the values derived from a stacked spectra of \simi150 LRDs (\citealt{PerezGonzalez-LRD2024}, P\'erez-Gonz\'alez et al. in prep.), the COSMOS-Webb median-stacked LRD \citep{Akins+25} and the unique LRD known as 'The Cliff' \citep{deGraaff+25}, respectively. In the left and middle panels, the brown diamond lie outside the colour range displayed. Black arrows in the lower right corners represent the vector magnitude for $A_\mathrm{V}$\eqq0.3\,mag.}
         \label{fig:color}
\end{figure*}

The large spectro-photometric dataset available for GN-z11 combining NIRCam, NIRSpec, MIRI, and MRS allows us to trace both the nebular lines and the continuum from rest- UV up to 0.86\micron. By removing the nebular line contributions from the aperture photometry, we obtained continuum-only fluxes, which allowed us to trace the GN-z11 underlying continuum emission (see Sect.~\ref{subsec:Cont-only}).

The continuum-only results show (see Table~\ref{tab:fluxes}) how, although the MIRI/F560W value agree within the uncertainties with the continuum level expected from previous SED fittings based on UV-only data \citep{Jiang+21, Tacchella+23_GNz11}, our F770W photometric value exceeds their predicted flux. In particular, once we take into account the difference with the photometry from \citet{Tacchella+23_GNz11} (see App.~\ref{app:Phot_norm}), our F770W continuum level (212\pmm20\nJy) is $>$70$\%$ larger ($>$5$\sigma$) than the SED fitting expected value when using UV-only photometry (\simi120\nJy). This excess at \simi0.66\micron rest-frame could be hinting at the presence of a non-negligible old stellar population or a red component that would be strongly emitting at optical-red and near-IR wavelengths. 

\begin{figure*}
\centering
   \includegraphics[width=\linewidth]{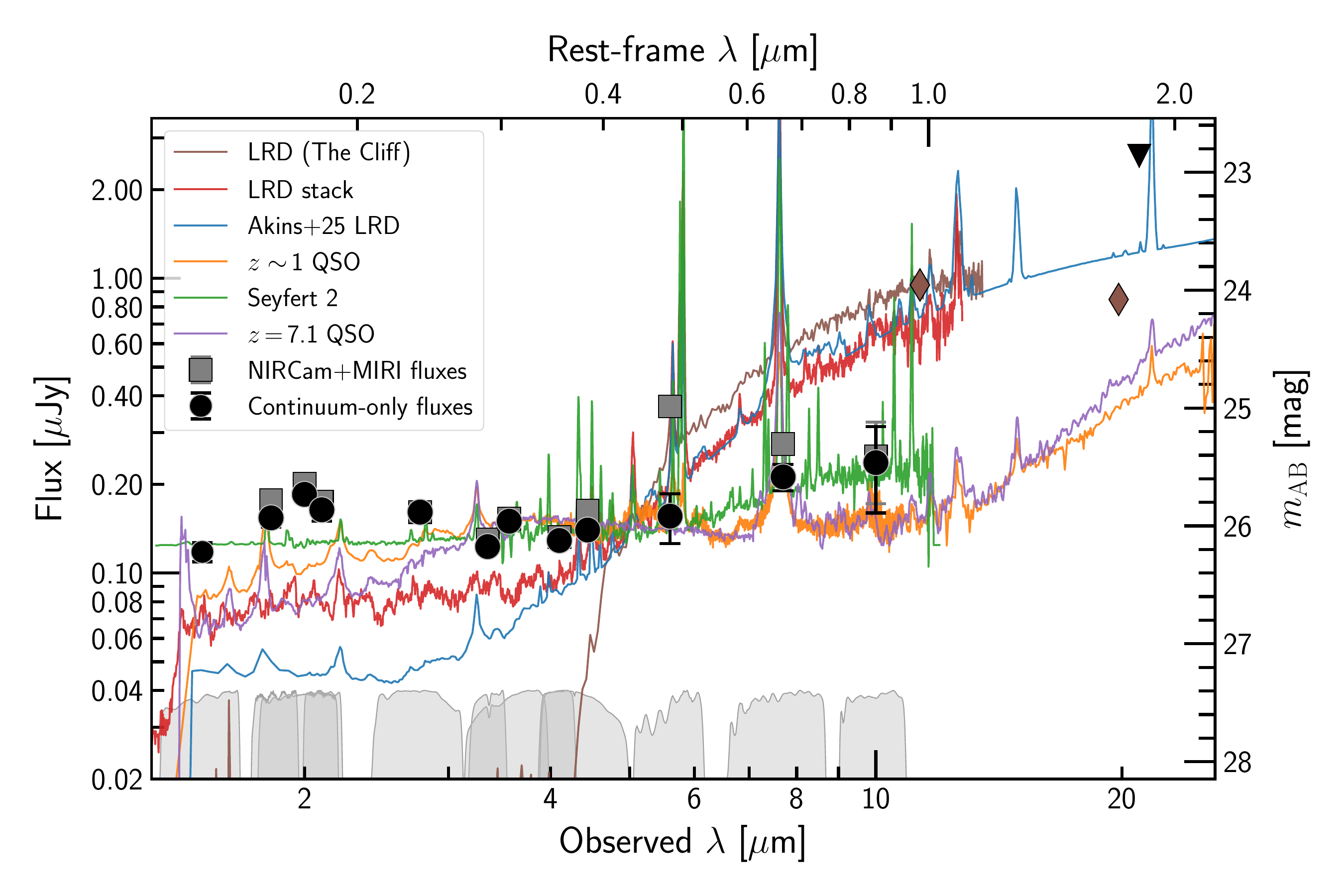}
      \caption{NIRCam and MIRI photometry compared with different red SED objects. Grey squares and black circles represent the NIRCam+MIRI observed and continuum-only (see Sect.~\ref{subsec:Cont.-results}) fluxes, respectively. Coloured lines represent the SEDs for different AGNs types and high-\z objects with optical-red colours presented in Sect.~\ref{subsec:Cont.-results}. The red line represents the stacked spectra for \simi150 local and high-\z LRDs (\citealt{PerezGonzalez-LRD2024}, P\'erez-Gonz\'alez et al. in prep.). The brown line displays the spectra from the exotic \z\simi3.5 LRD The Cliff, while brown diamonds represent its re-normalised fluxes from \citet{deGraaff+25} at rest- 0.9 and 1.9\micron. The blue line shows the COSMOS-Webb median-stacked LRD SED model presented in \citet{Akins+25}. The orange and purple lines show the \z\simi1 QSO composite spectrum, created by combining the UV+optical and near-IR spectral templates from \citet{Selsing+16} and \citet{Glikman+06}, and the NIRSpec+MRS spectrum for the \z{}\eqq7.1 QSO J1120+0641 (\citealt{Bosman2024,Bosman+25}), respectively. The HST/STIS spectrum for the Sy2 Mrk3 presented by \citet{Spinelli+06} is displayed as a green line. All observed spectra have been redshifted to \z{}\eqq10.6 and normalised at 4800\AA.}
         \label{fig:SED_templates}
\end{figure*}

A first step to investigate the nature of GN-z11, and this possible red excess, is to use classic colour-colour diagrams covering from rest-frame UV (0.17\micron) to optical-red (0.66\range0.86\micron) wavelengths. Figure~\ref{fig:color} displays the diagrams for the UV (F200W\range F444W), optical (F444W\range F560W), and optical-red (F560W\range F770W and F770W\range F1000W) colours derived from the continuum-only fluxes. In this figure we compare these continuum-only colours to those expected from stellar populations models and those from intrinsically red objects (e.g. QSO, AGNs, and LRDs). In particular, we considered the stellar population templates from BPASSv2.2 for continuum and instantaneous SFHs and a range of stellar ages limited to the age of the Universe at \z{}\eqq10.6 (i.e. 440\Myr). Their associated nebular continua were also included using the corresponding Cloudy models for those ages and SFHs \citep{Ferland+17}. For completeness, we displayed the age evolution tracks for metallicities 0.05 and 0.2\Zsun. Green diamonds represent the colours of the nuclear ultraviolet-optical spectrum of Mrk\,3. This low-\z galaxy is classified as a heavily obscured Seyfert 2 nucleus \citep{Guainazzi+16}, with a hidden BLR identified by the presence of broad Balmer lines in polarised light \citep{Miller1990}. Its spectrum, obtained with the HST/STIS spectrograph using an aperture size of 0.2\arcsec (\ie50\,pc at the distance of Mrk3, \citealt{Spinelli+06}), represents the SED of a luminous type 2 AGN, while minimising the potential contamination due to the stellar light that could be affecting larger apertures. Moreover, its line ratios (O3H$\gamma$\eqq$-0.1$, Ne3O2\eqq$-0.2$, and O32\eqq0.9), agree with those found in high-\z AGNs \citep{Mazzolari+24,Mazzolari+25}. Orange diamonds correspond to the colours measured for the template of luminous blue QSOs at intermediate redshifts (1\,<\,\z{}\,<\,2) constructed from the simultaneous ultraviolet to near-infrared spectroscopy \citep{Glikman+06,Selsing+16}. Purple diamonds represent the colours derived for the NIRSpec+MRS spectrum of the \z{}\eqq7.1 QSO J1120+0641, whose BLR and dust emission has been found to be consistent with those of lower-redshift quasars \citep{Bosman2024,Bosman+25}. We filled the gap between the NIRSpec and MRS data (\ie 3.12\range4.9\micron) following its best-fit SED shape ($f_{\lambda}$\,$\propto$\,$\lambda^{-0.25}$; \citealt{Bosman+25}). Brown diamonds show the colour derived from the spectra of the singular little red dot (LRD) presented by \citet{deGraaff+25}, known as `The Cliff'. This object presents the strongest Balmer break among LRDs and shows \Ha absorption, which are explained under the ‘black hole star’ (BH*) scenario in which dense gas surrounds a powerful ionising source (see also \citealt{Naidu+25} and \citealt{Taylor+25}). Finally, the red and blue diamonds display the colour derived for an updated stacked spectra from \citet{PerezGonzalez-LRD2024}, combining $>$150 nearby and high-\z LRDs, and the COSMOS-Webb median-stacked LRD from \citet{Akins+25}, respectively. The spectra of all these sources together with the GN-z11 JWST photometry are presented in Figure~\ref{fig:SED_templates}.

The continuum-only fluxes show an SED blue in the rest-UV (F200W\range F444W\eqq$-$0.30\pmm0.05\,mag) with a flattening in $f_\nu$ up to rest-frame 0.5\micron (F444W\range F560W\eqq0.12\pmm0.21\,mag). Beyond \OIII{$\lambda$5008}, the SED appears to rise, with colours F560W\range F770W\eqq0.33\pmm0.24\,mag and F770W\range F1000W\eqq0.18\pmm0.40\,mag (0.26\pmm0.24 and 0.33\pmm0.40\,mag, respectively, if we consider the impact of the potential broad hydrogen Balmer lines components, see Sect.~\ref{subsec:Cont-only}). Although the rest- UV and optical colours can be reproduced by a young (\simi10\Myr) stellar burst, it fails to explain the optical-red colours (\eg\,F560W\range F770W and F770W\range F1000W, see Figure~\ref{fig:color}). On the other hand, the Seyfert 2 spectrum can produce the optical-red continuum-only colours observed with MIRI but fails to explain the UV-blue colour F200W\range F444W. Finally, the LRD stacked and The Cliff spectra are too red, especially beyond 0.4\micron, and they are only marginally consistent with F770W\range F1000W due to the large uncertainty in the F1000W flux. These results, based on the traditional colour-colour diagrams, suggest that while the UV emission from GN-z11 can be produced by a young star formation burst, the presence of an extra red source is needed to explain its rest-frame optical-red colours. Note, however, that in these colour-colour diagrams, the stellar plus nebular continuum models do not contain any contribution due to small amounts of hot dust emitting at optical-red wavelengths. Likewise, mixed stellar populations that would produce optical colours different from those of (single age) instantaneous bursts are not represented.

In the following sections, we perform more detailed modelling using different SED fitting codes with different scenarios (\eg SFHs, AGN; see Sect.~\ref{subsec:SED_fitting}) to investigate the origin of these UV, optical, and optical-red colours. In addition, SED fitting also allows us to model both the continuum and the emission lines simultaneously, providing a more complete view of the physical mechanisms driving the emission in GN-z11.

\subsection{SF-only SED modelling}
\label{subsec:Onlystellar}

\begin{table*}
    \centering
\caption{Modelled fluxes derived from the SED fitting}
\begin{tabular}{l|c|ccccc}
\hline
 Line/Filter & Observed  & Single pop & Mixed pop. & Mix.+ty1 (CIG.) & Mix.+AGN (SYN.)  &  Mix.+ty2 (CIG.) \\
\hline
H$\beta$ & <44 & 22\pmm1 & 24\pmm0 & 22\pmm1  & 26\pmm4 & 22\pmm1 \\
\OIII{{}$\lambda$4960} & <44 & 44\pmm1 & 52\pmm1 & 44\pmm1  & 43\pmm10 & 46\pmm1 \\
\OIII{$\lambda$5008}& 136\pmm14 & 132\pmm4 & 156\pmm4 & 133\pmm3  & 128\pmm30 & 136\pmm1 \\
H$\alpha$ & 68\pmm9 & 63\pmm2 & 70\pmm2 & 61\pmm1 & 63\pmm10$^\dagger$ & 63\pmm1 \\
F560W & 368\pmm26 & 326\pmm4 & 382\pmm7 & 335\pmm5  & 357\pmm50 & 360\pmm2 \\
F770W & 274\pmm20 & 176\pmm1 & 202\pmm5 & 185\pmm5 &  261\pmm30 & 254\pmm5 \\
F1000W & 249\pmm77 & 111\pmm3 & 130\pmm6 & 126\pmm6 &  274\pmm40 & 285\pmm11 \\

\end{tabular}
    \label{tab:SED_fluxes}
    \tablefoot{The acronyms CIG. and SYN. stand for \texttt{CIGALE} and \texttt{SYNTHESIZER-AGN} (See Sects.~\ref{subsec:SED_fitting},~\ref{subsec:Onlystellar} and ~\ref{subsec:2comp+AGN}). Units are in 10$^{-19}$\ergscm and nanojanskys for the lines and photometric points, respectively. $^\dagger$: We derived a 1\EXP{-18}\ergscm flux for the broad component in \Ha.}
\end{table*}

\begin{table*}
    \centering
\caption{Physical parameters from SED fitting.}
\begin{tabular}{l|c|ccccc}
\hline
  & Observed  & Single pop & Mixed pop. & Mix.+ty1 (CIG.) & Mix.+AGN (SYN.)  & Mix.+ty2 (CIG.) \\
 \hline
sSFR [Gyr$^{-1}$] &  & 53\pmm6 & 33\pmm12 & 83\pmm20 & 85\pmm19 & 56\pmm4 \\
$M_\mathrm{*}$  [10$^{8}$M$_\odot$] &  & 5.6\pmm0.6 & 14.7\pmm1.2 & 3.7\pmm0.8 & 2.1\pmm0.9 & 5.3\pmm0.3 \\
$M_\mathrm{*}^\mathrm{young}$  [10$^{8}$\,M$_\odot$] &  & - & 1.8\pmm0.2 & 1.7\pmm0.2 & 0.5\pmm0.2 & 1.8\pmm0.1 \\
$t_\mathrm{young}$ [Myr] &  & - & 4\pmm1 & 6\pmm1 & 0.32\pmm0.15 & 6\pmm1 \\
$t_\mathrm{old}$ [Myr] &  & 21\pmm3 & 63\pmm23 & 39\pmm21 & 6\pmm2 &  20\pmm2 \\
%$t_\mathrm{all}$ [Myr] &  &  & &  & 3\pmm1 &   \\
SFR$_\mathrm{10Myr}$  [$M_\odot$\,yr$^{-1}$] & 24\pmm3$^\dagger$ & 30\pmm2 & 18\pmm3 & 18\pmm3 &  18\pmm5 & 18\pmm2 \\
%SFR$_\mathrm{100Myr}$  [$M_\odot$\,yr$^{-1}$] &  & 30\pmm2 & 34\pmm14 & 20\pmm9 & 30\pmm1 \\
log($L_\mathrm{AGN}$) [\ergs]  &  & - & - & 44.57\pmm0.04 &  44.6\pmm0.5 & 44.77\pmm0.02 \\
$A_\mathrm{V}$  [mag] & <0.2$^\dagger$ & 0.15\pmm0.01 & 0.13\pmm0.02 & 0.04\pmm0.01 & 0.10\pmm0.03 & 0.03\pmm0.01 \\
$Z^{\dagger\dagger}$ [\Zsun{}] & 0.17\pmm0.03$^\dagger$ & - & - & - & 0.16\pmm0.03 & - \\
EW(\Ha{}) [\Angs{}] & 547\pmm72 & 983\pmm55 & 932\pmm24 & 870\pmm38 & 540\pmm110 & 572\pmm17 \\
EW(\Hb{}) [\Angs{}] & 140\pmm19$^*$ & 175\pmm10 & 169\pmm5 & 166\pmm6 & 170\pmm40 & 146\pmm3 \\
EW(\OIII{$\lambda$5008}) [\Angs{}] & 786\pmm81 & 1144\pmm63 & 1159\pmm35 & 1080\pmm41  & 840\pmm230  & 932\pmm19 \\
\end{tabular}
    \label{tab:SED_results}
    \tablefoot{$^\dagger$: Values derived from MIRI/MRS data presented in \citet{Alvarez-marquez+25}.$^{\dagger\dagger}$: A fixed metallicity value was adopted during the \texttt{CIGALE} fits (see Sect.~\ref{subsec:SED_fitting}). $^*$: EW(\Hb{}) is derived from \Ha and F560W fluxes, assuming \Ha{}/\Hb\eqq2.86 (see Sect.~\ref{subsec:Cont-only}).} 
\end{table*}

\begin{figure*}
\centering
   \includegraphics[width=\linewidth]{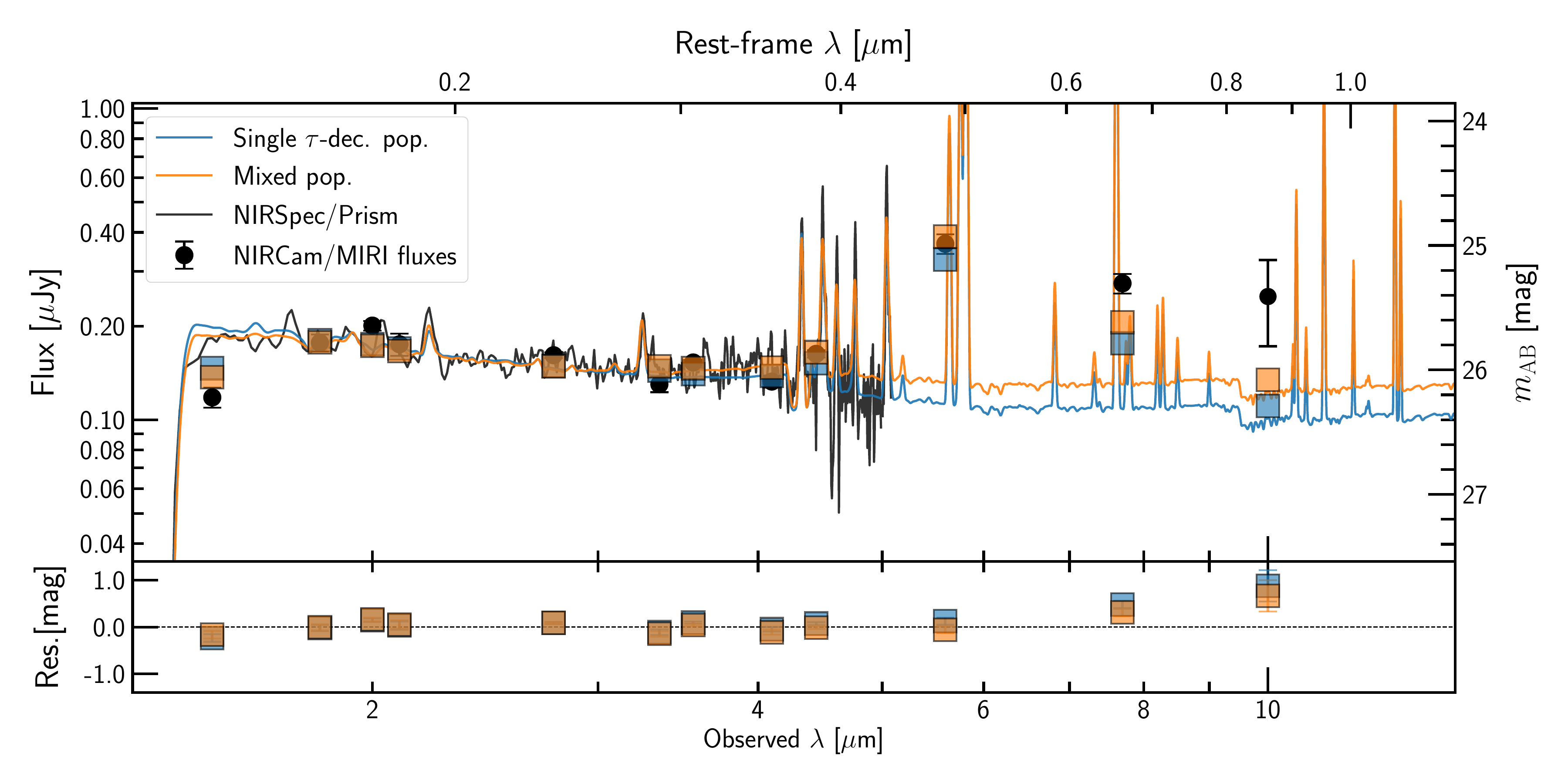}
      \caption{\texttt{CIGALE} SED models derived from the single and mixed stellar population scenarios described in Sect.~\ref{subsec:Onlystellar}. Blue (orange) line and squares represent the SED models and the expected fluxes for NIRCam/MIRI filters derived using a single $\tau$-decaying (mixed) stellar population. Black line and triangles represent the NIRSpec/Prism spectra and the NIRCam+MIRI photometry, respectively. The blue and orange lines have been convolved to match de NIRSpec/Prism resolution (R=100).}
         \label{fig:SED_young}
\end{figure*}

The colour-colour analysis carried out suggests an increase in the continuum emission beyond 0.66\micron. In this section we perform an SED analysis including both the emission lines and the photometric points to dig into the nature of this enhancement.

As a first approach for the \texttt{CIGALE} SED fitting, we considered that GN-z11 can be modelled by a stellar population following a tau-decaying SFH with $\tau$ varying in the [1, 5000\Myr{}] range, and with ages up to 440\,Myr (\ie age of Universe at \z\eqq10.603). Figure~\ref{fig:SED_young} shows the best-fit SED obtained from these adopted model parameters and the comparison with the observed NIRCam+MIRI fluxes. The Bayesian analysis results in an almost constant SFH ($\tau$\eqq2500\pmm1500\Myr) and a stellar population of \Mstar\eqq(5.6\pmm0.6)\EXP{8}\Msun and an age of $t$\eqq21\pmm3\Myr. Although for this model, we obtain \Ha and \OIII{$\lambda$5008} that are in agreement with those measured with the MRS, the modelled F770W flux (\ie176\pmm1\nJy) is underestimated by \simi45$\%$, 5$\sigma$ below the value observed with MIRI (see Table~\ref{tab:fluxes}). The blue optical continuum of this young stellar population makes this difference more pronounced in F1000W, modelling a flux \simi44$\%$ fainter (2$\sigma$ considering the large F1000W uncertainties). For completeness, we have tested an exponentially growing SFH (negative $\tau$), obtaining consistent results. 

Since a single young stellar population is not able to justify the UV and optical continuum and the nebular emission of this galaxy, we considered the possibility of having a mix of different stellar populations. In recent years, a number of works have proposed that stellar populations of different ages are required to fit the UV to optical SED of high-\z galaxies. \citet{Tacchella+23} found the presence of old stellar populations in \z>7 galaxies using SED fitting based on NIRSpec+NIRCam data. Underlying \simi100\Myr stellar populations have also been found in five 5<\z{}<9 galaxies in which the SEDs are dominated by blue emission produced in young bursty star-forming regions \citep{Gimenez-Arteaga+23}. Some authors have found that a double power law better describes the SFHs in high-\z galaxies, as single population models can be out-shone by young UV-bright stars, yielding to lower stellar masses \citep{Gimenez-Arteaga+24} . In addition, stellar populations older than \simi10\Myr produce a redder optical continuum, in accordance with the MIRI colours found in GN-z11 (see Fig.~\ref{fig:color}). 

Therefore, we also considered a mixed stellar component consisting of a young population ($t$\,$<$\,10\,Myr) with a constant star formation and a more mature one (20\,$<$\,$t$\,$<$\,440\,Myr) formed in a past burst. In this scenario, GN-z11 can be interpreted as a young compact burst in a mature host galaxy. Figure~\ref{fig:SED_young} displays the best-fit SED obtained under this approach and the residuals with the observed NIRCam+MIRI fluxes. Under this mixed stellar population scenario, GN-z11 is composed of a young (4\pmm1\Myr) stellar population that represents 12$\%$ of the total stellar mass (1.5\pmm0.1\EXP{9}\Msun), while the more mature one has an average age of 63\pmm23\Myr. These values are consistent with the double-component SED fitting carried out in \citet{Tacchella+23_GNz11}, in which the authors found a point-source with a young stellar population (\simi11\,Myr) along with a more massive and mature one (\simi10$^9$\Msun and \simi35\,Myr) dominating the extended emission. Although the near-IR emission produced by the mature stellar population increases the modelled flux in F770W up to 202\pmm5\nJy, it is still 3$\sigma$ fainter than the measured value (see Table~\ref{tab:SED_fluxes}). Even if we add the contribution of the upper limit of the potential broad component of \Ha ($<$2\range3\EXP{18}\,erg\,s$^{-1}$\,cm$^{-2}$, \citealt{Alvarez-marquez+25}), the flux in F770W would increase only by 18\range27\nJy and would be still incompatible with the MIRI data by almost 2$\sigma$ (see Table~\ref{tab:SED_fluxes}). Despite the increase in the modelled flux in F1000W (\ie130\pmm6\nJy) driven by the optical and near-IR emission from the older stellar population, we do not recover the observed MIRI flux (\ie249\pmm77\nJy).

Although assuming either a single or mixed stellar population provides a good SED fit up to rest-frame 0.5\micron, both approaches fail to reproduce the F770W and F1000W photometry. As is discussed in Sect.~\ref{subsec:Cont.-results}, the significant increase in the SED of GN-z11 beyond 0.66\micron cannot be reproduced solely by a combination of different stellar populations. This result suggests the presence of an additional, intrinsically red, component in GN-z11. 
%In the next subsection we modify the SED fitting to include sources of different nature to address discrepancies in the reddest MIRI bands.

\subsection{SF + AGN SED modelling}
\label{subsec:2comp+AGN}

\begin{figure*}
\centering
   \includegraphics[width=0.95\linewidth]{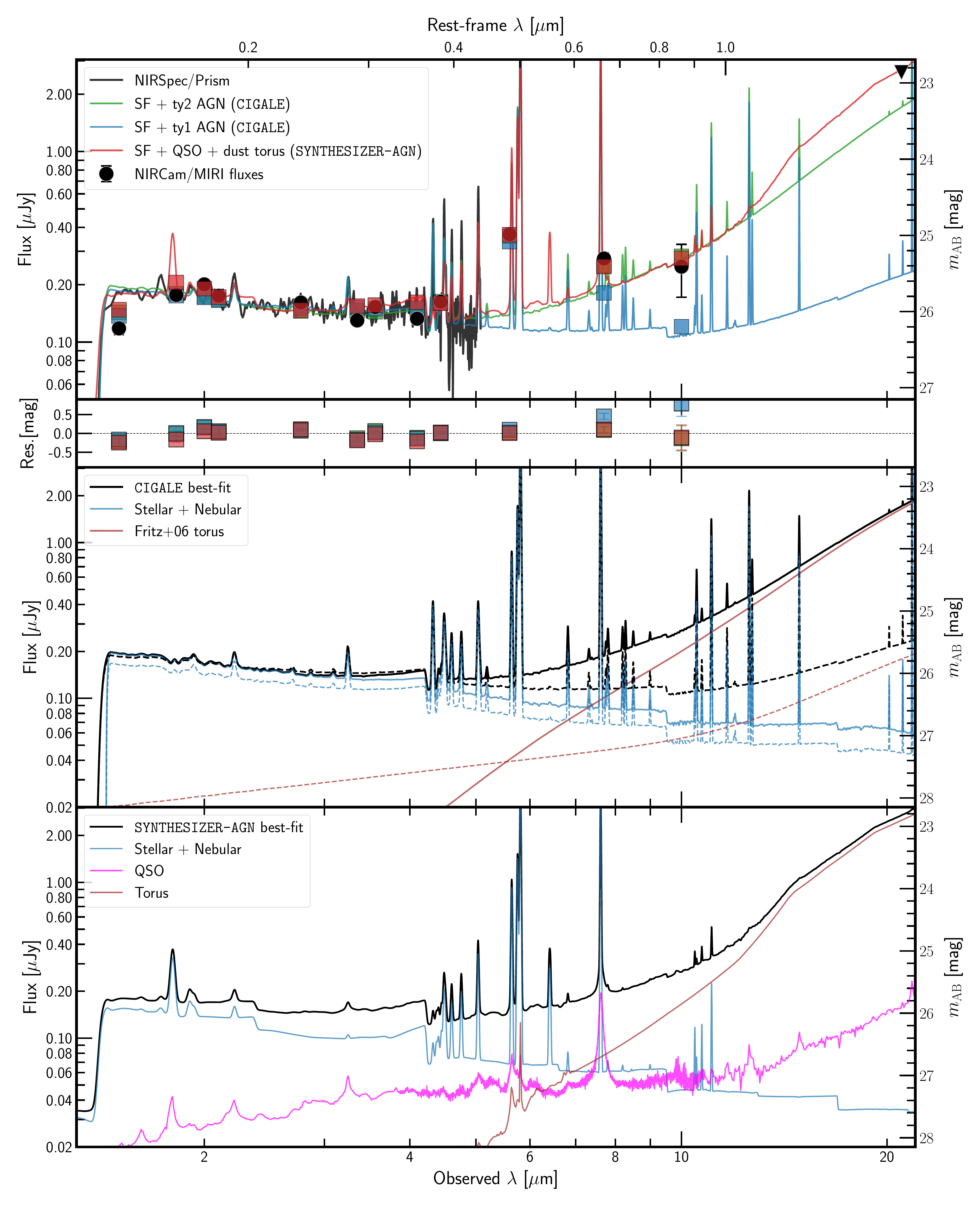}
      \caption{SED best-fit models considering the star formation and the different AGN scenarios. The top panel shows the best-fit SEDs obtained with \texttt{CIGALE} assuming type 1 and type 2 AGNs as blue and green lines, respectively. The red line displays the best-fit derived with \texttt{SYNTHESIZER-AGN} using a QSO template and an additional dusty torus. The black line and black circles represent the NIRSpec/Prism data presented in \citet{Bunker+23} and the NIRCam and MIRI photometry as derived in Sect.~\ref{subsec:phot}, respectively. Coloured squares show the modelled NIRCam and MIRI fluxes from the different best-fit models, while their residuals with the measured photometry are displayed below. Green, blue, and red lines have been convolved to match the spectral resolution from NIRSpec/Prism (\ie R=100). The middle panel shows the \texttt{CIGALE} best-fit models (in black) and the contributions of the nebular and stellar components (in blue) and the \citet{Fritz+06} torus emission (in brown). Dashed and continuous lines differentiate the results drawn from the SED fitting assuming a type 1 and type 2 AGN, respectively. The bottom panel displays the \texttt{SYNTHESIZER-AGN} best-fit (in black) along with the contributions of the nebular and stellar components (in blue), the QSO template (in magenta), and the dusty torus (in brown). }
         \label{fig:SED_AGNs}
\end{figure*}

In addition to the young starburst \citep{Bunker+23, Tacchella+23_GNz11, Alvarez-marquez+25}, some studies indicates the presence of high critical density lines and broad emission lines in the ultraviolet suggesting that GN-z11 could host a type 1 active black hole \citep{Maiolino+24_BH}.
To evaluate the presence of a potential AGN in GN-z11, we also considered the combination of star formation and an AGN (Type 1 or 2). The potential presence of an AGN would naturally produce a red component associated with the hot dust emission from the torus around an accreting black hole. The best-fit SED models for the different SF+AGN scenarios are presented in the top panel of Fig.~\ref{fig:SED_AGNs}. Specifically, \texttt{CIGALE} results including type 1 and 2 AGNs are represented in blue and green, respectively, while \texttt{SYNTHESIZER-AGN} result is displayed as a red line. Tables~\ref{tab:SED_fluxes} and ~\ref{tab:SED_results} summarise the photometric and physical results of these models.

Adding a type 1 AGN to the mixed stellar population modelled by \texttt{CIGALE} does not significantly improve the SED fitting beyond 0.66\micron. The top and middle panels from Fig.~\ref{fig:SED_AGNs} show how the type 1 AGN starts to dominate the emission only above 0.9\micron and the best-fit modelled F770W and F1000W fluxes are well below the observed values by factors 1.5 and 2, respectively. The type 1 AGN is modelled in \texttt{SYNTHESIZER-AGN} by a QSO composite template (see Sect.~\ref{subsec:SED_fitting}) that includes the contribution from broad emission lines and the dust emission in the optical-red and near-infrared spectral range. The top and bottom panels in Fig.~\ref{fig:SED_AGNs} show that the QSO template is almost flat until rest-frame \simi1\micron and it is not able to reproduce the observed fluxes in the MIRI reddest filters. Only when including an ad hoc extra dusty torus (\ie highly obscured QSO, see Sect.~\ref{subsec:SED_fitting}) are the modelled F770W and F1000W fluxes (257 and 258\nJy, respectively) compatible with the observed ones. However, the addition of this extra hot dust emission to the empirical spectrum of the QSO is not physically justified. As is shown in Figure~\ref{fig:SED_templates}, the $>$0.7\micron emission from the empirical \z\simi1 QSO SED agrees with the spectrum of the \z{}\eqq7.1 luminous QSO J1120+0641, for which JWST/MIRI spectroscopy has revealed a dust emission in the dusty torus with a temperature of about 1400\,K \citep{Bosman2024}. This agreement in the SED between the low-intermediate-\z and high-\z QSOs makes the physical justification for the presence of such an extra dusty torus emission difficult.

We consider next the presence of a type 2 AGN, whose UV and optical emission is expected to be obscured by the torus. According to the \texttt{CIGALE} best-fit model, the steep optical and near-IR continuum from the type 2 AGN emission increases the SED modelled fluxes up to 254\nJy for the F770W filter, being compatible with the MIRI measured fluxes (see Table~\ref{tab:SED_fluxes}). Despite the F1000W photometry not being taken into account during the SED fitting due to its low significance, the type 2 scenario yields a F1000W flux (285\nJy) compatible within the uncertainties, whereas the SED best-fit lies below the upper limit for F2100W. Although the available AGN models in \texttt{CIGALE} do not include the contribution from the emission lines produced by the AGN, the analysis carried out indicates that assuming a type 2 AGN yields a better agreement with the observed MIRI fluxes (\ie\,optical to red continuum SED slope) than using different type 1 scenarios.

Regardless of the AGN type considered, we derived similar ages for the different stellar populations. Specifically, we estimated ages of \simi4\range6 and \simi20\range40\Myr for the young and mature stellar populations, respectively. Ages for the stellar populations were derived independently for the spatially resolved point-source and extended stellar components (\simi 11 and 35\Myr, respectively) identified in NIRCam images \citep{Tacchella+23_GNz11}. Although the resolution of the MIRI images (\ie FWHM\eqq0.27\arcs for F770W) does not allow us to spatially resolve these two components (rest-UV $r_\mathrm{reff}$\eqq0.05\arcs for the extended one), our derived ages are compatible with the previous NIRCam photometry.

To ensure the robustness of these results, we replicated the \texttt{CIGALE} SED fits using the SKIRTOR AGN models \citep{Stalevski+12,Stalevski+16}, which include a clumpy dust distribution for the torus. These AGN models yield similar results, whereby only a type 2 AGN can explain the F770W and F1000W fluxes. In the next section, we discuss the general physical scenario drawn from this SED best-fit along with other non-AGN possibilities.

\subsection{Origin of the red excess emission}
\label{subsec:comparison_templates}

As has already been shown in the previous sections, the flux in the F770W filter, and in the low-significance (3$\sigma$) F1000W value, shows an excess that cannot be fully explained by the combination of young and intermediate-age stellar populations, even when including the contribution by the nebular emission (\ie the narrow \Ha line detected with the MRS). This excess hints at the presence of a source radiating at (rest-) wavelengths of 0.66 to 0.86\micron. The excess at these wavelengths can be interpreted as the emission produced by either the presence of a dusty torus of an AGN or dust associated with the extremely compact starburst identified in GN-z11.

\subsubsection{Dust in AGN: Dusty torus emission}
\label{subsub:AGN}

Although GN-z11 has not been detected in X-rays,  a 3$\sigma$ upper limit of $L_\mathrm{X}$(2$-$10\,keV)\,<\,3\EXP{43}\ergs has been derived assuming the typical photon index for narrow-line Seyfert 1 galaxies \citep{Maiolino+24_BH}. In addition, a value of $<$2.9\EXP{43}\ergs, based on its \Ha luminosity \citep{Ho+01,Alvarez-marquez+25}, could also be consistent with hosting a type 2 AGN. These upper limits agree well with the predicted  $L_\mathrm{X}$(2$-$10\,keV) based on the SED-fitting assuming the presence of type 1 and 2 AGNs. Using the X-ray extension of the CIGALE models (X-CIGALE, \citealt{Yang+20}), we obtain $L_\mathrm{X}$(2$-$10\,keV)\eqq(0.9\pmm0.1) and (2.1\pmm0.3)\EXP{43}\ergs for the best-fit SED considering the type 1 and 2 AGNs, respectively. Similar X-ray 2$-$10 keV luminosities (\ie 1.8\EXP{43} and 2.6\EXP{43}\ergs, respectively) were obtained when applying the $L_\mathrm{X}$(2$-$10\,keV)$-L_\mathrm{bol}$ relation from \citet{Marconi+04} and the bolometric luminosity derived from the SED fitting (see Table~\ref{tab:SED_results}). In summary, the predicted X-ray luminosities of the best-fit stellar plus AGN (both type 1 and 2) SEDs are compatible with the non-detection of X-ray emission in GN-z11.

However, the scenario of a type 1 AGN is not favoured by the present results. The potential presence of an undetected, weak broad H$\alpha$ line of 2\range3\EXP{-18}\ergscm tracing a type 1 AGN (\citealt{Alvarez-marquez+25}) would contribute to about 18\range27 nJy to the F770W flux, which is not sufficient to explain the excess of 91\nJy found during the SED fitting (see Sect.~\ref{subsec:2comp+AGN}). In addition, Figure~\ref{fig:SED_templates} displays how luminous intermediate \z\simi1 \citep{Glikman+06,Selsing+16} and \z$>$7 QSOs \citep{Bosman2024} have a SED at rest-wavelengths 0.6\range0.8\micron flatter than otherwise detected at F770W and F1000W in GN-z11. The SEDs of these QSOs are consistent with hot dust from the black hole surrounding torus emitting at a temperature of about 1400\,K, contributing at wavelengths of 1\micron and above (\eg\citealt{Bosman2024}). Moreover, our \texttt{SYNTHESIZER-AGN} best SED fit combining the stellar populations and type 1 AGN requires an additional ad hoc red source (see Sect~\ref{subsec:2comp+AGN}) in addition to the hot (\simi1300\range1400\,K) dust component to explain the observed flux excess at 0.66 and 0.86\micron. As was already mentioned in the previous section, this extra component is not physically motivated, and therefore the scenario of a standard type 1 AGN is not supported by the present data.

Under the scenario of a Seyfert 2 contribution, the AGN bolometric luminosity derived by \texttt{CIGALE} can be used to estimate the size and mass of the dusty torus assuming that its physical parameters (\ie the radial extent and inclination of the torus, the number of clouds along the equator of the torus, the angular and radial distribution of clouds, and the optical depth per cloud) are similar to those derived for low-\z Seyfert 2 galaxies. Following the clumpy dusty torus \citep{Nenkova+08a,Nenkova+08b,Garcia-Bernete+19}, the inner radius of the torus ($R_\mathrm{sub}$) is given as
\begin{equation}
 R_\mathrm{sub}[\mathrm{pc}] = 0.4\times \left(\frac{1500\,\mathrm{K}}{T_\mathrm{sub}}\right)^{2.6}\times \left( \frac{L_\mathrm{AGN}}{10^{45}\mathrm{erg\,s^{-1}}} \right)^{0.5}
 ,\end{equation}
 
where $L_\mathrm{AGN}$ is the AGN bolometric luminosity 
and $T_\mathrm{sub}$ is the dust sublimation temperature. Assuming the radial extent of the torus ($R_\mathrm{T}$) to be about 10\range30 times $R_\mathrm{sub}$ and a dust sublimation temperature of 1500 K, we obtain that the radius of the dusty torus would be in the 3\range9\,pc range for the derived AGN luminosity of 5.9\EXP{44}\ergs (see Table~\ref{tab:SED_results}). Following the $M_\mathrm{dust}$ expression by \citet{Garcia-Bernete+19} and assuming the typical number of clouds, their opacity, and their radial and angular distribution in Seyfert 2 galaxies (\ie\,$N_0$\eqq14, $\tau_V$\eqq70, $q$\eqq0 and $\sigma$\eqq60\degr, respectively), we derive a torus mass of 0.5\range1.5\EXP{5}\Msun. Thus, the size and mass of the potential torus in GN-z11 would be in the range of those detected in low-\z Seyfert 2 galaxies with median size and mass of 3.5\pmm3.9\,pc and (3.9\pmm5.1)\EXP{5}\Msun, respectively \citep{Garcia-Bernete+19}. Although the SED fitting and colour analysis of the red excess are compatible with the presence of a type 2 AGN, the true nature of this potential AGN along with its resemblance to low-\z counterparts are beyond the scope of this work and need to await the confirmation of the 10 micron flux and the extension of the photometry into redder wavelengths with additional MIRI imaging.

\subsubsection{Dust emission in compact starbursts}
\label{subsub:starburst}

The high SFR and stellar mass along with the compact size of GN-z11 indicate that this galaxy has extreme stellar mass and star formation surface densities ($\Sigma_\mathrm{SFR}$\,=\,9.33\,$\times$\,10$^2$\,M$_{\odot}$\,yr$^{-1}$\,kpc$^{-2}$, \citealt{Alvarez-marquez+25}). In addition, the UV and optical line ratios have shown that this target is a low-metallicity galaxy with low dust extinction \citep{Bunker+23, Alvarez-marquez+25}. This is in agreement with scenarios of UV bright galaxies at \z{}>10 in which the narrow-line component of the nebular emission is barely extinct as the dust would be evacuated outside these extreme compact starbursts by the outflows \citep{Fiore+23, Ziparo+23, Dekel2023, Li&Dekel+24}. 
In fact, recent studies have identified the existence of dusty outflows in RXCJ2248-ID (\z{}\eqq6.1; \citealt{CrespoGomez+25_RXC}) and J0217-0208 (\z{}\eqq6.2; \citealt{Marques-Chaves+25}) as was predicted in the Attenuation-Free model (AFM, \citealt{Ferrara+23,Fiore+23}. Similarly to GN-z11, these galaxies are also identified as N-emitters \citep{Topping+24,Marques-Chaves+25}. These two galaxies show large extinctions in the outflowing ionised gas traced by the broad emission line components identified in their optical spectra. Extinctions of $A_\mathrm{V}$\eqq1.5\range2.5 magnitudes are measured, while the blue compact starburst appears almost extinction-free as traced by the narrow Balmer line components and the rest-UV continuum emission. These results show that, despite a low dust-extinction in the UV-bright starburst nucleus, a considerable amount of dust is present in these compact N-emitters. The detection of similar dusty outflows in GN-z11 would require extremely deep spectroscopy with MIRI in order to detect the presence of broad line components in the optical hydrogen and oxygen lines.

Moreover, complex dust and gas structures in the interstellar medium around young compact massive clusters have been detected in low-metallicity galaxies. Recent NIRCam and MIRI imaging of the low-\z galaxy IZw18 \citep{Hirschauer+24} has shown in detail the complexity of the dusty star formation in this extremely metal-poor galaxy (which has a gas-phase oxygen abundance of 3\% solar, see \citealt{Hirschauer+24} and references therein). The presence of massive evolved stars, dust-enshrouded stellar objects, and shells and filaments of dust and/or gas due to supernovae has been identified. As is shown in the images of IZw18 presented in \citet{Hirschauer+24}, the dust in and around the stellar clusters could be heated by the strong radiation field, re-radiating in the optical-red and near-infrared wavelength range, and resulting in a flux excess above the stellar+nebular continuum at these wavelengths. Flux excess at wavelengths of 0.66\range0.86\micron as measured in GN-z11 has already been claimed in the low-\z metal-poor galaxies SBS0335-052E \citep{Reines+08, Adamo+10_SBS} and Haro11 \citep{Adamo+10_Haro11}. Notably, \citet{Adamo+10_SBS} found that many of the young and massive stellar clusters present in SBS0335-052E show flux excesses amounting to 0.2 to 0.5 magnitudes above their predicted stellar continuum at rest-frame \simi0.8\micron. These galaxies have a hard radiation field ($>$\,54.4\,eV), as is demonstrated by the presence of strong \ion{He}{II}\,1640 and \ion{He}{II}\,4686 emission lines \citep{Kehrig+18, Sirressi+22, Wofford+21}, and even more recently demonstrated by the detection with MIRI spectroscopy of extended mid-infrared [\ion{Ne}{V}]\,14.32\micron line emission, with an ionisation potential of 97\,eV, in SBS0335-052E \citep{Mingozzi+25} and IZw18 \citep{Arroyo-Polonio+25}.

A photoluminiscence mechanism (Extended Red Emission, or ERE, see \citealt{Witt+04}) has been hypothesised to explain this excess at 0.7\range0.9\micron in these low-metallicity galaxies (e.g. \citealt{Reines+08, Adamo+10_SBS}). This non-thermal mechanism originates from UV-excited small dust grains or complex molecules in dense circumstellar environments. However, if the excess flux observed in MIRI extends beyond one micron, it would not be consistent with the ERE scenario. If the SED continues to rise, the most likely explanation would be thermal emission of dust close to the sublimation temperature (\ie 1000\range1500\,K). Although uncommon, some dense nucleus and young clusters in low-metallicity galaxies show near-IR excess compatible with the presence of dust emitting at 600\range1100\,K (Mrk996, \citealt{Thuan+08}; NGC5253, \citealt{Vanzi+04,Calzetti+15}). The presence of $>$1000\,K dust has also been observed around WR stars \citep{Marchenko+07,Lau+23}, whose existence has been proposed as a potential origin of the unusual large N/O ratio in GN-z11 \citep{Cameron+23,Gunawardhana+25}. 
%On the other hand, if the SED flattens out beyond \simi1\micron, it would resemble that observed in LRDs and LRD-BH* objects, as showed in Figure~\ref{fig:SED_templates}. 

Recent NIRSpec and MIRI spectroscopy has clearly detected PAH emission, including the strong 3.3$\mu$m feature, in the dominant super-cluster (SSC-N) of the nearby  galaxy IIZw40 \citep{Lai2025}. This indicates that dust and complex molecules are produced and survive in low-metallicity (0.1\range0.2\,$Z_{\odot}$, \citealt{Leitherer2018}) extreme radiation environments such as those present in SSC-N, with a radiation field ($G$) 8.6\EXP{4} times higher than the Habing radiation field ($G_0$\eqq1.6\EXP{-3}\ergscm, \citealt{Habing1968}). The radiation field in GN-z11, derived from the extreme UV luminosity ($L_\mathrm{1500}$\eqq 1.7$\times$10$^{29}$ erg s$^{-1}$\,Hz$^{-1}$, \citealt{Bunker+23}) and small size of the galaxy ($r_\mathrm{eff}$\eqq64\,pc, \citealt{Tacchella+23_GNz11}), is a factor of 5 higher (\ie $G/G_0$\eqq4.3\EXP{5}) than that of SSC-N. Therefore, dust and complex molecules could also be present in the low-metallicity (0.17\,$Z_{\odot}$, \citealt{Alvarez-marquez+25}) interstellar medium of GN-z11 as in IZw40. The presence and emission processes of dust in very extreme radiation environments such as those present in GN-z11 have to be explored further with MIRI imaging at longer wavelengths covering the near-IR rest-frame.

\section{Summary and conclusions}
\label{sec:Summary}

This paper presents new imaging of GN-z11 with the MIRI F560W, F770W, and F1000W filters, which extend the rest-frame UV-optical coverage up to 0.86\micron. In these MIRI images, GN-z11 shows a compact structure, compatible with previous results based on NIRCam imaging \citep{Tacchella+23_GNz11}. We have combined these new MIRI images with archival NIRSpec/Prism and MIRI/MRS spectroscopy, and NIRCam imaging, to analyse the SED of GN-z11 covering the rest-frame 0.12\range0.86\micron. In addition, the completeness of the spectroscopic NIRSpec and MRS dataset allowed us to remove the emission line contribution from the NIRCam and MIRI broad-band photometry, deriving the SED of GN-z11 with line-free, continuum-only fluxes. The main results drawn from this analysis can be summarised as follows:

\begin{itemize}

\item Classical colour-colour diagrams have been used to compare the observed colours with several SED templates including stellar populations of different ages, AGNs, and high-\z LRDs. None of the templates can explain the colours of GN-z11 over the entire spectral range covered by our data. The rest-frame UV-optical colours (i.e. F200W\range F444W  versus F444W\range F560W) of GN-z11 are compatible with the SEDs of unobscured young and intermediate age instantaneous bursts (including nebular emission), while not consistent with their optical and optical-red colours  (i.e. F444W\range F560W, F560W\range F770W and F770W\range F1000W). The SEDs of high-\z LRDs and QSOs are not consistent with any of the UV-optical-red colours. The spectrum of Mrk\,3, classified as a classical Seyfert 2, is compatible with the optical-red colours but fails to explain the observed UV-optical colours.

\item The GN-z11 SED beyond \simi0.66\micron cannot be entirely explained by the stellar and associated nebular emission alone. The observed flux in the F770W filter (rest-frame 0.66$\mu$m) shows an excess of 36\pmm3\% above the predicted flux for a mixed stellar population of young (4 Myr) and intermediate-age (63 Myr) stars. The flux marginally detected in the F1000W filter (rest-frame 0.86\micron) represents an even larger excess, although one with larger uncertainties (\ie\,91\pmm28\%).

\item Both the continuum-only colours and the SED analysis covering the entire 0.12\range0.86\micron spectral range identify a optical-red continuum excess at wavelengths 0.66\range0.86\micron. This hints at a red source that appears undetected in the ultraviolet and represents a minor contribution to the emission in the optical range (\ie$<$0.6\micron). Although the origin of this excess is unclear, we propose that it could be produced by the emission from the dusty torus of a type 2 AGN characterised by a radius of 3\range9\,pc with a mass of (0.5\range1.5)\EXP{5}\Msun. Alternatively, the excess could be related to hot dust emission or photoluminiscence of dust (Extended Red Emission) in the presence of the extreme radiation field (G/G$_0$= 4.3$\times$10$^5$) produced by the compact and massive starburst in GN-z11, as has already been hypothesised for some low-\z low-metallicity starbursts. The presence of a standard type 1 AGN is not supported by the present data as an additional ad hoc red source is required to explain the observed flux excess. 

\end{itemize}

In summary, flux excess has been identified in the MIRI/F770W (rest- 0.66\micron) and also, although with lower significance, in MIRI/F1000W (rest- 0.86\micron) images of GN-z11. This excess is likely due to the dust emission associated with an AGN dusty torus or with dust in the environment of an extremely compact and young starburst. However, the current wavelength coverage is still limited and the true nature of the heating source is still unclear.
Deep imaging with MIRI/F1000W and redder filters are required to confirm the 0.86\micron detection and extend the SED coverage beyond \Ha, characterising its near-IR spectral shape up to at least 1.1\micron. Imaging with deep integrations using MIRI F1000W and F1280W filters will be executed with JWST in the near future (PID 7605). This will help to discriminate between the two proposed scenarios. In addition, deeper MIRI/MRS spectroscopy than is in existence would  also be required to detect or rule out the presence of a weak broad H$\alpha$ associated with a \simi10$^6$\Msun type 1 AGN, or identify the existence of dusty outflows.

\begin{acknowledgements}
A.C.G. acknowledges support by JWST contract B0215/JWST-GO-02926.
L.C. and J.A-M. acknowledge support by grants PIB2021-127718NB-I00 and PID2024-158856NA-I00, A.A-H. by grant PID2021-124665NB-I00, and P.G.P-G by grant PID2022-139567NB-I00 funded by the Spanish Ministry of Science and Innovation and the State Agency of Research MCIN/AEI/10.13039/501100011033 and ERDF "A way of making Europe". 
M.A. acknowledges financial support from Comunidad de Madrid under Atracción de Talento grant 2020-T2/TIC-19971.
D.L. was supported by research grants (VIL16599, VIL54489) from VILLUM FONDEN.
A.B. and G.O. acknowledge support from the Swedish National Space Administration (SNSA). 
AJB acknowledges funding from the "FirstGalaxies" Advanced Grant from the European Research Council (ERC) under the European Union’s Horizon 2020 research and innovation programme (Grant agreement No. 789056).
L.A.B. acknowledges support from the Dutch Research Council (NWO) under grant VI.Veni.242.055 (\url{https://doi.org/10.61686/LAJVP77714}) and the ERC Consolidator grant 101088676 ("VOYAJ").

This work is based on observations made with the NASA/ESA/CSA James Webb Space Telescope. The data were obtained from the Mikulski Archive for Space Telescopes at the Space Telescope Science Institute, which is operated by the Association of Universities for Research in Astronomy, Inc., under NASA contract NAS 5-03127 for \textit{JWST}; and from the \href{https://jwst.esac.esa.int/archive/}{European \textit{JWST} archive (e\textit{JWST})} operated by the ESDC.
\end{acknowledgements}

\bibliographystyle{aa} 
\bibliography{bibliography.bib}

\begin{appendix}

\section{NIRCam-NIRSpec-MIRI normalisation}
\label{app:Phot_norm}

As mentioned in Sect.~\ref{subsec:phot}, we used the NIRCam photometry to re-normalise the NIRSpec spectra. In this appendix we introduce the main steps followed during this re-normalisation.

First, we PSF-matched the NIRCam images to the MIRI/F560W resolution. This is done by deriving empirical PSFs (ePSFs) based on the stars available in the FoV (\ie between 3 and 6 non-saturated in each filter). We then create the corresponding kernel for each NIRCam image using the MIRI PSF models from \citet{Libralato+24}, as there are not enough available stars in the MIRI FoV. Once the convolution has been applied, we extracted the NIRCam and MIRI photometry using circular apertures of $r$\eqq0.5\arcsec (see Sect.~\ref{fig:Aperture_plot} for details). As a consistency check, we extracted the photometry at the native spatial resolution of each filter, applying the corresponding aperture corrections. The resulting fluxes are consistent with those obtained from the PSF-matched data within uncertainties.

We noticed that our NIRCam photometry is slightly larger (20\range25$\%$) than the one derived in \citealt{Tacchella+23_GNz11} for a smaller aperture ($r$\eqq0.35\arcsec). In our analysis, we have used a more recent data release from JADES (\ie\,DR3) where the NIRCam images were reduced with improved calibration files from JWST. In addition, we have used a larger and more distant from GN-z11 annular aperture to derive the background level. The combined effect of using a more distant aperture to derive the background level and more recent calibration files and pipeline favour the detection of diffuse emission that was missed in the original calibrated images.

Finally, we convolved the NIRSpec/Prism spectra with the NIRCam filter transmission curves to derive the expected fluxes in each filter. These NIRSpec continuum values were therefore compared with the NIRCam photometry, measuring a median 32.5$\%$ offset that remains nearly constant with wavelength, showing variations of less than 5$\%$. We had also compared the NIRCam photometry with the NIRSpec pseudo-continuum fluxes obtained in 750\Angs bins (see Sect.~\ref{subsec:SED_fitting}), obtaining similar results. We therefore multiply the NIRSpec/Prism spectra by 1.325 to match the NIRCam and NIRSpec data. \citet{Bunker+23} reported an agreement between the NIRSpec/Prism reduction, which included slit-loss corrections for point-like source, and the $r$\eqq0.1\arcsec{} NIRCam photometry from \citealt{Tacchella+23_GNz11}. This difference is produced by as GN-z11 UV morphology deviates from an ideal point-like source, which is assumed during the data reduction. In fact, even larger differences between NIRCam and NIRSpec/MSA (\ie 30\range50$\%$) have been found in other high-\z galaxies (\eg \,JADES-GS-z14-0, \citealt{Carniani+24}).

For consistency, we tested whether this difference in the UV flux would lead to any noticeable difference during the SED fitting. Therefore, we replicated the SED analysis described in Sect.~\ref{subsec:2comp+AGN}, with a double stellar population and the presence of an AGN, considering the NIRSpec pseudo-filters prior normalisation. As in this work we assumed a uniform re-normalisation, most of the properties derived with the UV slope remains unaltered. However, having a lower UV continuum results in a slightly older (\ie 6\Myr vs 4\Myr) and more massive young stellar population (\ie 13$\%$ larger). As the \Ha and the optical and near-IR fluxes do not change, we do not observe significant differences either in the SFR, the total stellar mass or the AGN luminosity.

\end{appendix}

\end{document}